\documentclass[conference]{IEEEtran}
\IEEEoverridecommandlockouts
\usepackage{cite}
\usepackage{amsmath,amssymb,amsfonts}
\usepackage{algorithmic}
\usepackage{graphicx}
\usepackage{textcomp}
\usepackage{hyperref}
\usepackage{xcolor}
\def\BibTeX{{\rm B\kern-.05em{\sc i\kern-.025em b}\kern-.08em
    T\kern-.1667em\lower.7ex\hbox{E}\kern-.125emX}}
\begin{document}

\title{Koedds: A National Real Estate Investment Analysis
}

\author{\IEEEauthorblockN{Sean Kouma}
\IEEEauthorblockA{\textit{Computer Science Department} \\
\textit{Colorado State University}\\
Fort Collins, Colorado \\
skouma@colostate.edu}
\and
\IEEEauthorblockN{William Edds}
\IEEEauthorblockA{\textit{Computer Science Department} \\
\textit{Colorado State University}\\
Fort Collins, Colorado \\
wre136@colostate.edu}
}

\maketitle

\begin{abstract}
With costs and risks increasing for investors and home buyers alike, additional analysis of the housing market is required to help individuals make the right choice. In addition to traditional market analysis, other aspects such as the economic vulnerabilities of the local community must be taken into account to further ensure real estate buyers receive a positive return on investment from their purchases as well as ensuring that traditional home owners get the best price for their future home. 
\end{abstract}


\section{Introduction}
During the COVID-19 pandemic, real-estate markets saw an initial drop in all forms of activity within the United States. However, in less than a few months, this trend was reversed and one of the largest investments in residential real estate began to occur. This surge in the residential real estate market has led to low inventory for homes as well as high prices for potential buyers. Reasons for this sudden surge include record-low mortgage rates, personal needs to upgrade to larger spaces and the need for increasing ones own wealth by means of flipping houses or by generating passive income by way of renting to tenants~\cite{BankrateCovidHousingMarket}. With these reasons exemplifying why the real-estate market accelerated during and after COVID-19, it was alarming to find that recently, investors are now beginning to pull out of the market. 2023 saw a drastic decline in investments~\cite{RedfinHousingPullOut}. It is suspected that this decline in investor activity could be caused by over saturation of the market, rising interest rates on home loans, and investment in poorly analyzed opportunities. For investors, this leads to issues of not being able to find tenants to fill vacant properties or not being able to offer the rentals at a price that community members can afford. For traditional home owners, immediate issues relate to the purchase cost of a home being unprecedentedly high, to the point where new home owners to median income couples are not able to afford such property. Essentially, average Americans are being priced out of the housing market. 

To be able to combat these potential road blocks, home owners and investors need to have a deeper understanding of both the housing market as well as other factors that can influence the decision making process. Doing so will enable individuals to make more informed decisions which could result in a lower price for the purchase of a home, or result in more consistent tenant occupation of a rental property.

Fortunately, there is no shortage of data on real estate along with a number of other economical and social factors that frequently influence real estate markets. The goal set out for this project was to use this abundantly available data on historical real estate growth and combine it in a scoring model with the CDC's Social Vulnerability Index~\cite{SVI}. The result will be a score which can be used to compare counties for the potential real estate growth while minimizing associated risks.

\section{Problem Characterization}
Deciding where to purchase a home is a large hurdle for both investors and residents alike. For the vast majority of home owners in the United States, a majority of their personal wealth and net worth is contained within the value of their home and any other real estate they may own\cite{USPortfolio}. And while Mark Twain might be right in saying to "Buy land, they're not making it anymore"\cite{MarkTwain} it is still not clear exactly where one should purchase land. Conventional wisdom says that the best time to buy real estate was ten years ago but that the second best time to buy real estate is now. That gives us an ideal time to buy land, but should we purchase real estate in our hometown? Someplace that has shown strong historical growth? An area that is very economically stable? Ideally maybe we want all of the above but it is difficult to achieve all three. The problem that we desire to solve is to provide a solution to that problem. After all, a well made decision can set up families to pass on some generational wealth for one or more generations and could be the difference between a parent leaving their child with nothing but debts to pay versus leaving them with a wealth that will be useful. This example might be somewhat contrived considering a majority of individuals don't receive a significant inheritance\cite{inheritance} it still serves to show the importance of making good decisions when purchasing real estate.

Ultimately, the problem here is clear. We want clear guidance on where to invest our money in real estate such that it appreciates as much as possible while incurring as little risk as possible. We also want it to perform well during times of economic downturns and other times of volatility. Our last big requirement, is that this guidance should be data-driven based off of historical performance and available risk factors.

\section{Dominant Approaches to the Problem}
Fortunately for us, trying to determine real estate trends is nothing new. One approach that partially inspired this research was the development of a systematic housing crash index by Daniel Reynolds, Joseph Riva, and Greyson Sequino in the spring of 2022\cite{HCI}. Their approach consisted of bringing in a variety of factors to score counties on how vulnerable the county was to potentially experiencing a real estate market crash. They brought in an abundant amount of data ranging from wage data, the Gini income inequality index, the ratio of income to poverty level, and occupancy rates. They combined all of this data into a single risk score to assign the county. In a lot of ways, this approach was the inverse of our project. While their paper analyzes county risk, we are focused on potential upsides for counties.

Another aspect of analyzing real estate trends involves looking at the effects of pandemics on real estate markets. In the 2021 paper "COVID-19's Impact on real estate markets: review and outlook", authors Balemi, Fuss, and Welgand discuss the effects of pandemic events on real estate trends. Previous pandemics such as cholera outbreaks in London in the 19th century as well as the SARS outbreak in Hong Kong in 2003 led to sharp declines in housing demand, increasing supply and lowering sales prices.~\cite{Balemi2021}. This is further stated as the authors discuss similar results following an increase in cases of leukemia found in Nevada in 2004.

Balemi and team further evaluated the housing market in response to COVID-19. They observed an initial dip in home sale transactions at the start of the pandemic in the United States in March of 2020. However, unlike previous events, transactions grew by more than double that amount by August of 2020, leading to the highest average home value the United States had ever seen according to the Housing Market Index by the National Association of Home Builders. Evidence for the increase in transactions stems from situations such as employees taking advantage of telework-compatible jobs, home owners being able to adjust to larger homes due to the necessity of sustaining more individuals working and studying from home, and individuals being able to afford those new homes with record low interest rates. While it is impossible to predict when global pandemics or other volatile times will take place, we do want some way to account for them when thinking about real estate purchases. \cite{Balemi2021}

\section{Methodology}
After determining the problem we were trying to solve and looking at neighboring approaches, we needed to collect our data, process our data to get some results, and finally turn those results into something that could easily be visualized and used to guide decisions. At a high level our plan involved combining historical real estate growth data at the county level with some additional data that would give us an idea of how risky investing in certain areas was. We then wanted to combine all of those data points in a weighted average that would represent the investment and purchase potential of buying real property in specific counties.

\subsection{Acquiring of Data}
For real estate growth data, one of the best metrics is using median sales price per county and measuring the change from month to month over a ten to fifteen year period. While Zillow has some data on current median sales prices along with current home valuations\cite{Zillow}, they seem to lack the long term data that was desired. Luckily, we came across Redfin, a technology-driven real estate brokerage which freely shares large datasets containing monthly real estate metrics\cite{Redfin}. This data is available going all the way back to 2012 when Redfin began collecting their data. The datasets also consist of metrics from the United States at the national, metro, state, county, city, zipcode, and even neighborhood level. The list of metrics recorded range from the beginning to the end of the recorded time period, region, median sales price, homes sold, pending sales, inventory amounts, new listings, price drops, and a lot more.

Our other major source of data was the social vulnerability index (also known as SVI) available from both the CDC directly or from urban-sustain.org. This index represents a vulnerability index for a county which effectively represents how resilient or vulnerable a county is if some sort of disaster were to take place. It could be an economic disaster, a pandemic like we saw in 2020, or a physical disaster like an earthquake or a hurricane. The index is composed of 16 social factors which includes poverty, lack of vehicle access, and crowded housing among others. While four different SVI sub-indexes are available including socioeconomic status, household characteristics, racial and ethnic minority status, and housing type/transportation, we pulled and used the combined score for our analysis.

\subsection{Consolidation of Collected Data}
Given the vast amount of data provided in the Redfin dataset, which was close to one gigabyte, the first step in the analysis was to eliminate unneeded data and narrow the results to valid entries. Using PySpark with a Spark cluster set up with ten machines, all but a few fields were removed from all records. Fields that were left were: period start, period end, region, property type, homes sold month-over-month, homes sold year-over-year, median sales price month-over-month, and median sales price year-over-year. Additionally, records missing this data were removed from the dataset so that no errors would occur in our calculations.

The remaining data was sorted by region and then by period end date. Using the month-over-month and year-over-year fields, averages for homes sold and median sale price per region were calculated and added to the dataset. Using these values, the range for averages was remapped so that they were normalized between zero and one, as to better display counties that saw the greatest increase in sales verses counties that have the least growth or greatest loss.

The social vulnerability index was already usefully normalized between zero and one, so no need to do anything there.

\subsection{Creating a Weighted Score}
Once we had all of our aggregated data points it was time to join the data by county and turn it into a weighted average. We developed a Python script to match the counties based off of name and joined the data into a single pandas dataframe. Once both pieces of information were available for every county it was a piece of cake to use pandas to merge and then normalize the values. One interesting item is that a low SVI score is desirable while a high sale price score is ideal. Thus we normalized the sale price score with a \[1 - SVI_{county}\] score to ensure our scoring model stayed consistent.

\subsection{Visualization of Data in Monocle}
After scoring all of the counties based off of our investment model, the next step involved visualizing our results. The first step of this process involved converting the two letter state codes to their associated FIPS codes. FIPS stands for Federal Information Processing Standards and is an agreed upon way to identify states with numbers from one to fifty. Since we did not want to reinvent the wheel to create a GeoJSON file with all of the shapes for every county, instead we used a file we found online that already had all of the shape data and we just added some additional fields to each county. This file used state FIPS codes along with the names of the counties after dropping 'County' or 'Parish' at the end of the county name. Ultimately we ended up using a regular expression with the FIPS code and county name to match the associated county in the file, and then added the relevant investment score field to that county. After doing all of this, we finally had a GeoJSON file with all of the data that we had generated thus far.

Our final step involved using the Monocle service available via Urban Sustain\cite{US}. The Monocle service is a data visualization tool that accepts a GeoJSON file upload and provides the ability to visualize the data overlaid on a colorful map. Monocle visualizes values at the high end of any given domain as red and the low end as blue, so we inverted our scores for the sake of the visualization such that a score near one was bad and would show as red and a 'good' investment score would show as blue. After making these changes and uploading the file we ended up with the following.

\begin{figure}[htp]
    \centering
    \includegraphics[width=9cm]{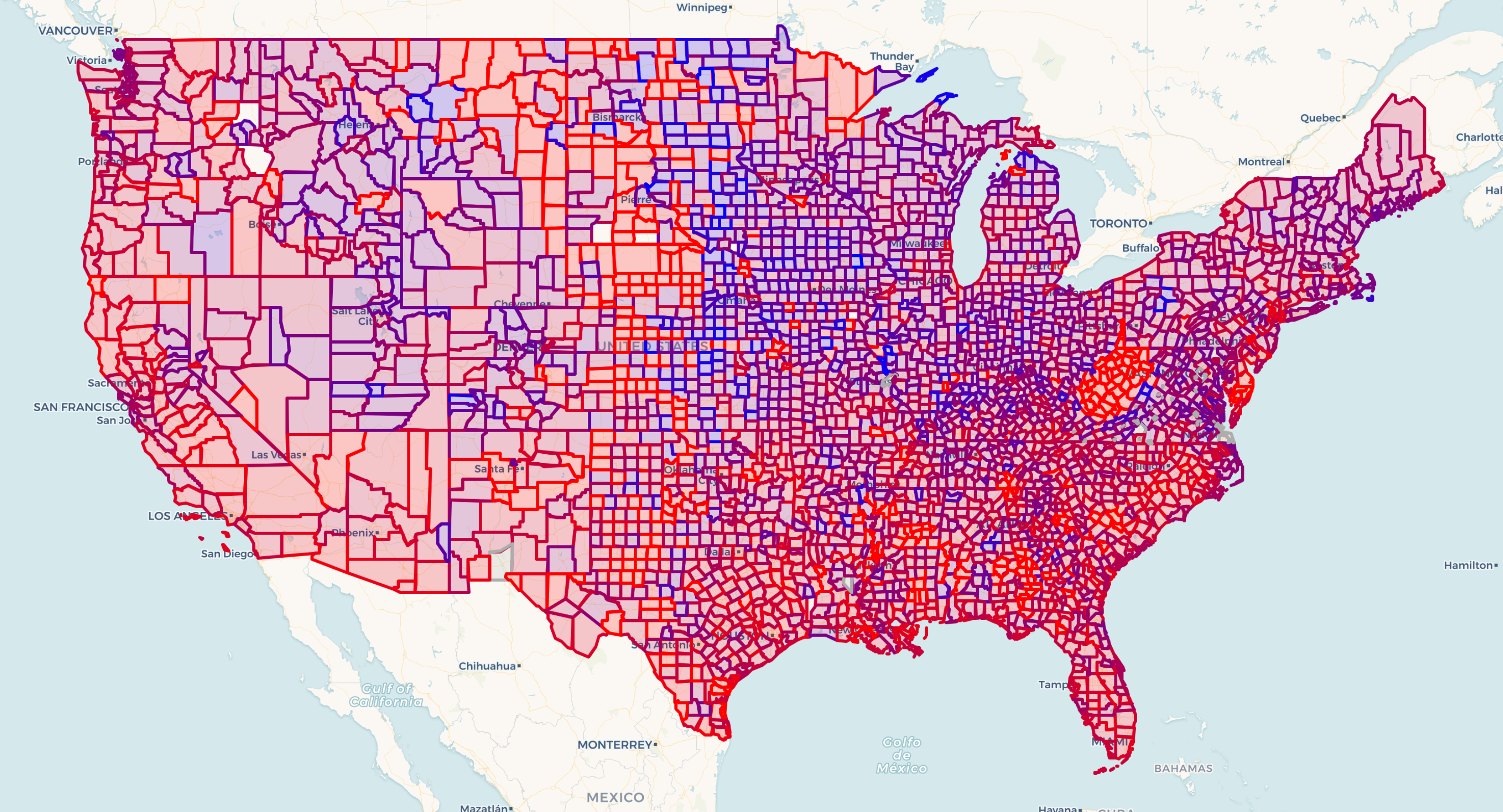}
    \caption{Investment Score Visualization}
    \label{fig:investment}
\end{figure}

\section{Experimental Benchmarks}
Experimental benchmarks is an area in which our research did poorly. We had difficulty coming up with a set of benchmarks that holistically captured our model's likely performance going forwards. For example, we could have just re-run our analysis going up until 2022 or 2023 and compared our recommended counties with the actual growth that those counties saw in the past year or perhaps year-to-date but this analysis is missing part of our goal. Thinking logically, a large portion of our model is attempting to take into account resiliency against potential real estate market downturns. We did not see this very much in the past year and a half and so it would mean relatively little whether our model performed well in these conditions.

One experimental benchmark that we did do was to calculate the correlation coefficient between the SVI score and the average annual real estate growth a county experienced. We hypothesized that there would be a statistically significant positive correlation between the two. After all, if the SVI score is an indicator of how resilient or vulnerable the county is, we should see that reflected in growth levels. Unfortunately for us, this did not end up being the case. After calculating the correlation coefficient it came out to be 0.126. While this is a slight correlation, any coefficient less that 0.2 is generally considered to be little to no correlation and not statistically significant. While we were disappointed with these results, there are still some insights to glean from it which are outlined in the following section. 

\section{Insights Gleaned}
A number of insights were gleaned over the course of this project. First of all, as mentioned above, we realized that there is not a strong correlation between real estate market growth and a county's economic stability as represented by SVI scores. This was rather surprising and points to there being some factors that we did not take into account. On the one hand, SVI is a combination of 16 different factors, several of which deal with minority groups which (however unfortunate it may be) likely don't play a significant role in real estate markets and can even have a negative impact depending on the part of the country the county is in. A second explanation might be that there \textit{is} a correlation for pretty economically stable counties but in the case of less stable counties there is more of a high-risk, high-reward situation taking place. Perhaps averaging those opposites hides the underlying factors. At any rate, these are guesses and are little more than speculation on why our original hypothesis was incorrect.

Another major insight is that after taking a look at the visualization we see that counties in the midwest, specifically those located around Iowa and Ohio seem to be ripe for investment based off our scoring model. This came as a bit of a surprise as we had anticipated that some of the fastest growing states like Colorado, Texas, and Arizona would have been scored the highest. Granted, most of the areas in red still have average annual growth levels in the 8 - 15\% range so they are by no means bad for investment, but they were still outperformed by more Midwestern regions. Perhaps the social vulnerability index played a large role in those areas performing well as they aren't known for being especially vulnerable.

\section{Proposed Problem Space Evolution}
Significant growth in this problem space going forwards seems to be almost certain. Machine learning continues to experience significant hype and as new and improved AI models come out at a staggering pace, we anticipate that going forwards a majority of real estate growth and investment models will primarily rely on machine learning. This is already visible in several places but perhaps most notably as Zillow regularly releases market forecasts based off their own Zillow Home Value Index \cite{Zillow}. With that in mind, we would guess that compound models which combine multiple different factors (as shown in this paper combining county resiliency and historical growth) to produce investment recommendations will likely be created by companies seeking to make a profit and will attempt to limit their model's public exposure as much as possible. After all, if a company \textit{really} has a winning model for where to invest their funds it is not in their best interest to share that with other potential competitors. We also would not be surprised if some of the more advanced models going forwards comes from private equity firms like BlackRock and Vanguard. This past year there has been some significant news relating to how much real estate is already owned by private equity firms and so due to their immense resources they have the most to gain from improving their investment portfolio.

Despite the corporate investment we expect to see in this space, it is also our hope that companies like Zillow will continue to produce forecasts that can be easily used by consumers. According to the U.S. Census Bureau only 65.9\% of homes are currently owned by their occupants\cite{USCB}. This leaves some area for significant growth, but as previously covered, sky high prices have made it difficult for a lot of consumers to be able to own their homes. With that in mind, being a well informed consumer when purchasing a home is more important than ever and we applaud and look forward to organizations which help consumers to make more informed decisions.

\section{Conclusions}
In summary, we are excited to see how our model stacks up over the next five to ten years. While we do not anticipate our model being widely used, it is still a useful personal tool if nothing else and the underlying data points are useful to keep in mind when considering real estate purchases. Based off this work, if any family or friends ask about the best place to see their money grow, we will be recommending that real estate purchases in and around Iowa are a pretty safe bet as they have seen consistent historical returns and the risks of a large crash there are minimal. These same results could also be used to inform other styles of real estate investment such as property tax liens as well. The most surprising aspect of this work was certainly the lack of a correlation between price growth and economic stability and we posit that this area is deserving of some additional research.

One final note is that while our research sought to highlight the ripest regions of the country for growth and stability, there were really only a small handful of counties that actually had an average price change year over year that was slightly negative. Keep in mind that this is out of the 3,143 counties that are in the United States. The vast majority of the counties had average returns in the upper single digits. Thus, our biggest takeaway is simply to invest in real estate whenever possible - regardless of the location. It is an almost guaranteed good investment in the long term. This seems consistent with a quote by Andrew Carnegie who once said that "Ninety percent of all millionaires become so through owning real estate. More money has been made in real estate than in all industrial investments combined. The wise young man or wage earner of today invests his money in real estate".\cite{MarkTwain}

\bibliographystyle{plain} 

\begin{thebibliography}{99}

\bibitem{Balemi2021}
Balemi, Nadia, Roland Füss, and Alois Weigand. 
\newblock COVID-19's impact on real estate markets: review and outlook.
\newblock {\em Financial Markets and Portfolio Management}, 35(4):495–513, 2021.
\newblock DOI: \href{https://doi.org/10.1007/s11408-021-00384-6}{10.1007/s11408-021-00384-6}.

\bibitem{BankrateCovidHousingMarket}
Rae Hartley Beck.
\newblock How did COVID affect the housing market?
\newblock 2023.
\newblock Available at: \url{https://www.bankrate.com/real-estate/covid-impact-on-the-housing-market/}.

\bibitem{RedfinHousingPullOut}
Dana Anderson.
\newblock Real Estate Investors Pull Back, Buying 45\% Fewer Homes Than a Year Ago.
\newblock 2023.
\newblock Available at: \url{https://www.redfin.com/news/investor-home-purchases-drop-q2-2023/}.

\bibitem{SVI}
Centers for Disease Control and Prevention/ Agency for Toxic Substances and Disease Registry/ Geospatial Research, Analysis, and Services Program.
\newblock CDC/ATSDR Social Vulnerability Index.
\newblock 2020.
\newblock Available at: \url{https://www.atsdr.cdc.gov/placeandhealth/svi/data_documentation_download.html}.

\bibitem{USPortfolio}
John Bailey Jones, Urvi Neelakantan.
\newblock Portfolios Across the U.S. Wealth Distribution.
\newblock 2023.
\newblock Available at: \url{https://www.richmondfed.org/publications/research/economic_brief/2023/eb_23-39}.

\bibitem{MarkTwain}
Jessica Willens.
\newblock 30 Inspiring Real Estate Quotes That Will Change Your Life.
\newblock 2022.
\newblock Available at: \url{https://realwealth.com/learn/real-estate-quotes/}.

\bibitem{inheritance}
Eric Reed.
\newblock Average American Inheritance, By Wealth Level.
\newblock 2024.
\newblock Available at: \url{https://finance.yahoo.com/news/average-american-inheritance-wealth-level-130120356.html}.

\bibitem{HCI}
Daniel Reynolds, Joseph Riva, Greyson Sequino.
\newblock Housing Crash Index.
\newblock 2022.
\newblock Available at: \url{https://docs.google.com/document/d/19lec4f3smnJJFPI7wtthBmKaL0CeZ_ny7WutzzoxKlE/edit}.

\bibitem{Zillow}
Zillow.
\newblock Housing Data.
\newblock 2024.
\newblock Available at: \url{https://www.zillow.com/research/data/}.

\bibitem{Redfin}
Redfin.
\newblock Housing Market Data.
\newblock 2024.
\newblock Available at: \url{https://www.redfin.com/news/data-center/}.

\bibitem{US}
Urban Sustain Monocle Service.
\newblock 2024.
\newblock Available at: \url{https://urban-sustain.org/services/monocle/}.

\bibitem{USCB}
Tony Mariotti.
\newblock Homeownership Statistics (2024).
\newblock 2024.
\newblock Available at: \url{https://www.rubyhome.com/blog/homeownership-stats/}.

\end{thebibliography}

\end{document}